\documentclass[prl,twocolumn,notitlepage,superscriptaddress,nofootinbib,amssymb,aps]{revtex4-1}
\usepackage{enumerate}
\usepackage{bm}
\usepackage{bbm}
\usepackage{enumitem}
\usepackage[usenames,dvipsnames]{xcolor}
\usepackage{amsfonts}
\usepackage{amssymb}
\usepackage{graphicx}   
\usepackage{hhline}
\usepackage{subfig}
\usepackage{hyperref}
\usepackage{color}
\usepackage{verbatim}
\usepackage{amsmath,amsthm,amssymb}
\usepackage{float}
\usepackage{breqn}
\usepackage{graphics}
\usepackage{tikz}
\usepackage[utf8]{inputenc}
\usepackage{autobreak}
\usepackage{verbatim}
\usepackage{mathtools}
\usepackage{tensor}

\definecolor{hyperref}{RGB}{026,028,087}

\newcommand{\beq}{\begin{equation}}
	\newcommand{\eeq}{\end{equation}}
\newcommand{\bea}{\begin{eqnarray}}
	\newcommand{\eea}{\end{eqnarray}}
\def\be{\begin{equation}}
	\def\ee{\end{equation}}

\def\beq{\begin{equation}}
	\def\eeq{\end{equation}}

\renewcommand{\L}{\mathcal L}

\def\be{\begin{equation}}
	\def\ee{\end{equation}}
\def\ba{\begin{eqnarray}}
	\def\ea{\end{eqnarray}}

\usepackage[normalem]{ulem}

\def\ba{\begin{eqnarray}}
	\def\ea{\end{eqnarray}}

\def\L{\mathcal{L}}

\def\({\left(}
\def\){\right)}

\allowdisplaybreaks

\begin{document}

\preprint{Imperial/TP/2022/MC/02}	
	
\title{Cotton Double Copy for Gravitational Waves}

	\author{Mariana Carrillo González}
	\email{m.carrillo-gonzalez@imperial.ac.uk}
    \affiliation{Theoretical Physics$,$ Blackett Laboratory$,$ Imperial College$,$ London$,$ SW7 2AZ$,$ U.K.}
	\author{Arshia Momeni}
	\email{arshia.momeni17@imperial.ac.uk}
	\affiliation{Theoretical Physics$,$ Blackett Laboratory$,$ Imperial College$,$ London$,$ SW7 2AZ$,$ U.K.}
	\author{Justinas Rumbutis}
	\email{j.rumbutis18@imperial.ac.uk}
	\affiliation{Theoretical Physics$,$ Blackett Laboratory$,$ Imperial College$,$ London$,$ SW7 2AZ$,$ U.K.}

\begin{abstract}
We construct a double copy relation between the Cotton spinor and the dual field strength spinor of topologically massive theories, as the three-dimensional analogue of the Weyl double copy. The relationship holds in curved backgrounds for wave solutions. We give an explicit proof for Type N spacetimes and show examples satisfying the Cotton double copy.
\end{abstract}

\maketitle

\section{Introduction}
The non-linearities of gravitational theories lead to involved perturbative calculations and intriguing features of exact classical solutions. In recent years, it has become apparent that a relation dubbed the {\it double copy} \cite{Kawai:1985xq,Bern:2008qj, Bern:2010ue}, which allows us to write gravitational amplitudes as the {\it square} of Yang-Mills (YM) amplitudes, can shed light into fundamental properties of gravity while allowing us to perform computations in a simpler framework. It is now known that this relationship holds beyond the scattering amplitudes case and it can relate classical solutions \cite{Saotome:2012vy, Monteiro:2014cda, Luna:2015paa, Luna:2016due, White:2016jzc, Cardoso:2016amd, Luna:2016hge, Luna:2018dpt, Godazgar:2020zbv, Goldberger:2016iau, Goldberger:2017frp, Ridgway:2015fdl, De_Smet_2017, Bahjat_Abbas_2017, Carrillo-Gonzalez:2017iyj,Anastasiou:2014qba,Anastasiou:2016csv,Anastasiou:2017taf,Anastasiou:2017nsz,Anastasiou:2018rdx,Borsten:2015pla,Goldberger_2018, Li_2018, Lee_2018, Plefka_2019, Berman_2019, Kim:2019jwm, Goldberger:2019xef, Alawadhi:2019urr, Banerjee:2019saj, CarrilloGonzalez:2019gof,Shen:2018ebu,Chacon:2021wbr,Bahjat-Abbas:2020cyb,Alfonsi:2020lub,Luna:2020adi,White:2020sfn,Alkac:2021bav,Alawadhi:2020jrv,Huang:2019cja,Keeler:2020rcv,Elor:2020nqe,Farnsworth:2021wvs,Lescano:2021ooe,Ferrero:2020vww,Gumus:2020hbb,Guevara:2021yud,Pasterski:2020pdk,Chacon:2021hfe,Alawadhi:2021uie,Cho:2021nim, Gonzalez:2021ztm,Chacon:2021lox}. For example, in 4 dimensions (4d) there is an interesting double copy relation for plane waves at linearised order which gives the Weyl tensor as the square of the YM field strength
	\begin{equation}
		W^{\text{lin.}}_{\mu\nu\rho\lambda}=\frac{1}{2}\frac{F^{\text{lin.}}_{\mu\nu}F^{\text{lin.}}_{\rho\lambda}}{e^{i p\cdot x}} \ .
	\end{equation}
This relationship has been shown to hold for exact solutions of Petrov Type D and Type N spacetimes when written in terms of spinors \cite{Luna:2018dpt,Godazgar:2020zbv}, and it is known as the Weyl double copy. The Weyl double copy has recently been explained form a twistors perspective \cite{White:2020sfn,Chacon:2021wbr,Chacon:2021lox}, and it has also been formulated in tensorial form in \cite{Alawadhi:2019urr,Alawadhi:2020jrv}. 

An interesting question is whether the double copy holds beyond the massless case. As studied in \cite{Johnson:2020pny,Momeni:2020hmc,Momeni:2020vvr,Gonzalez:2021bes,Gonzalez:2021ztm,Moynihan:2020ejh,Moynihan:2021rwh,Burger:2021wss,Hang:2021fmp,Hang:2021oso,Gonzalez:2022mpa}, there are only a couple of known cases where the scattering amplitudes arising as the double copy of massive theories lead to a well-defined local theory. One example requires an infinite tower of massive states that satisfies a special relationship between the masses in the theory. The second example corresponds to topologically massive gauge theories in 3d and we will focus on this case for the rest of the letter. 

\section{Topologically Massive Theories}
We start by introducing the topologically massive theories that are related through the double copy. The action of Topologically Massive Yang-Mills (TMYM) in a curved spacetime is,
	\begin{align}\label{eq:tpYM}
	S_{TMYM}=\int d^3x\sqrt{-g}\Bigg(-\frac{1}{4}F^{a\mu\nu}F_{a\mu\nu}+\frac{g}{\sqrt{2}} A^{\mu a}J_{\mu a} \nonumber\\
	+\varepsilon_{\mu\nu\rho}\frac{m}{12}\left(6 A^{a\mu }\partial^{\nu}A^{\rho}_{a}+g\sqrt{2}f_{abc}A^{a\mu}A^{b\nu}A^{c\rho}\right)\Bigg),
	\end{align}
where $m$ is the mass of the gauge field, $g$ the coupling strength, and $\varepsilon_{\mu \nu \rho}$ is the Levi-Civita tensor given by $\varepsilon_{\mu \nu \rho}=\sqrt{-g} \epsilon_{\mu \nu \rho}$, with $\epsilon_{\mu \nu \rho}$ the Levi-Civita symbol. We will consider gauge fields and sources of the form $A^{\mu \ a}=c^a A^\mu$ and $J^{\mu \ a}=c^a J^\mu$, with $c^a$ a constant color charge, so that the equations of motion become linear and read
	\begin{equation}
		\nabla_\mu F^{\mu\nu}+\frac{m}{2}\varepsilon^{\nu\rho \gamma}F_{\rho\gamma}=\frac{g}{\sqrt{2}} J^\nu \label{eq:TMYM},
	\end{equation}
where $F_{\mu\nu}$ is the linearised Yang-Mills field strength. The double copy of TMYM corresponds to Topologically Massive Gravity (TMG) whose action is
    \begin{align}
    \label{eq:tpmGR}
	S_{TMG}=&\frac{1}{\kappa^2}\int d^3x\sqrt{-g}\Bigg(-R+2\Lambda+\L_{Matter} \nonumber\\
	&-\frac{1}{2m}\varepsilon^{\mu\nu\rho}\Big(\Gamma^{\alpha}_{\mu\sigma}\partial_{\nu}\Gamma^{\sigma}_{\alpha\rho}+\frac{2}{3}\Gamma^{\alpha}_{\mu\sigma}\Gamma^{\sigma}_{\nu\beta}\Gamma^{\beta}_{\rho\alpha}\Big)\Bigg) \ ,
	\end{align}
where $\kappa^2=16\pi G$, and $\Lambda$ is the cosmological constant. The equations of motion read
\begin{equation}
G_{\mu\nu}+C_{\mu\nu}/m=-\kappa^2 \frac{T_{\mu\nu}}{2}-\Lambda g_{\mu\nu} \ , \label{eq:tmg}
\end{equation}
where $G^{\mu\nu}$ is the Einstein tensor and $C^{\mu\nu}=\varepsilon^{\mu\alpha\beta}\nabla_{\alpha}(R^{\nu}_{\beta}-\frac{1}{4}g^{\nu}_{\beta}R)$ the Cotton tensor.

We proceed to understand whether one can construct an analogue of the Weyl double copy for topologically massive theories. When trying to generalise this to 3d, one immediately hits a roadblock since the Weyl tensor is zero. Instead, we will look at the analogue of the Weyl tensor in 3d which is the Cotton tensor. In 3d, the Cotton tensor is invariant under conformal transformations and thus is zero for conformally flat spacetimes, just like the Weyl tensor for $d>3$. Since the Cotton tensor appears in the TMG equations of motion, Eq.~\eqref{eq:tmg}, this tells us that we could write it as a {\it{square}} of terms in the TMYM equations of motion. By a simple counting of derivatives, we see that an appropriate ansatz is
	\begin{equation}
		C^{\text{lin.}}_{\mu\nu}=-\frac{1}{4}\frac{\left(\partial^\lambda F^{\text{lin.}}_{\lambda(\nu}\right)\Big(\varepsilon_{\mu)\rho \gamma}(F^{\rho\gamma})^{\text{lin.}}\Big)}{e^{i p\cdot x}} \ , \label{eq:motivationDC}
	\end{equation}
which is satisfied for plane waves. Note that we can use the TMYM equations of motion, \eqref{eq:TMYM}, to rewrite this relation in a simpler form. Considering only localised sources, outside of the source we have
    \begin{equation}
		C_{\mu\nu}^{\text{lin.}}=\frac{m}{2}\frac{{}^\star {F}_{(\mu}^{\text{lin.}}{}^\star {F}_{\nu)^{\text{lin.}}}}{e^{i p\cdot x}} \ ,
		\label{eq:CottonDClin}
	\end{equation}
where ${}^\star {F}_{\rho}=\varepsilon_{\mu\nu\rho}F^{\mu\nu}/2$ is the dual field strength. In the following section we will use this relation as motivation for constructing the analogue of the Weyl double copy for topologically massive theories.

\section{3d spinor formalism and the Cotton double copy}
The spinor formalism in 3d has been considered in \cite{Milson:2012ry,castillo20033,doi:10.1063/1.1592611}. It uses the fact that the 3d Lorenz group $SO(1,2)$ is isomorphic to $SL(2,\mathrm{R})/\mathrm{Z}_2$ to rewrite the tangent space Lorentz transformations. This allows us to write a vector in tangent space as $v_a=-(\sigma_a)_{AB}v^{AB}$, where the sigma matrices form a basis of $SL(2,\mathrm{R})$ and satisfy the Clifford algebra. To move between coordinate space and tangent space we use the frame $e^\mu_a$ that satisfies $\eta_{ab}={e^\mu}_a {e^\nu}_b g_{\mu\nu}$. Thus, we can write a vector in coordinate space as $v^\mu=-{e^\mu}_a(\sigma^a)_{AB}v^{AB}$. The indices $A, B={1,2}$ are lowered and raised with the 2d Levi-Civita symbol $\epsilon_{AB}$ according to the following conventions $	\psi^A=\psi_B \epsilon^{BA} \ , \ \psi_A= 	\epsilon_{AB} \psi^B $ \cite{castillo20033}. In the following, it will be useful to work with a spinor basis given by a dyad $(\iota,o)$ that satisfies $\iota_A\iota^A=o_A o^A=0 \ ,\  \iota_A o^A=-1$. Thus we can write $\epsilon_{AB}=2\iota_{[A}o_{B]}$.

We note that in a vacuum 3d spacetime, the TMG equations of motion together with the definition of the Cotton tensor and the Bianchi identities tell us that
	\begin{equation} \label{eq:C eom}
   \nabla^{EA}C_{BECD}=\frac{m}{\sqrt{2}}{C^A}_{CDE} \ , 
	\end{equation}
where $C^{\mu\nu}=\sigma^\mu_{AB}\sigma^\mu_{CD}C_{ABCD}$. Meanwhile, the equation of motion for linearised TMYM and the Bianchi identity for the field strength give
	\begin{equation} \label{eq:f eom}
	   \nabla^{CA}f_{BC}=\frac{m}{\sqrt{2}} {f^A}_B \ ,
	\end{equation}
where the dual field strength is given by $f^\mu=-\sigma^\mu_{AB}f^{AB}$. Motivated by the linear relationship found in Eq.\eqref{eq:CottonDClin}, we propose that the analogue of the Weyl double copy between the Cotton and field strength spinors is \footnote{Note that the mass factor in this relation is a choice of conventions to match our motivation from Eq.\eqref{eq:motivationDC}. It could be absorbed in $S$ or we could write the relation for the traceless Ricci spinor since the TMG equations tell us that $C_{ABCD}=-m \Phi_{ABCD}$, where the traceless Ricci tensor is given as $S^{\mu\nu}\equiv R^{\mu\nu}-R g^{\mu\nu}/3=\sigma^\mu_{AB}\sigma^\mu_{CD}\Phi_{ABCD}$.}
    \begin{equation} \label{eq:cottonDCspinor}
		\boxed{
	    C_{ABCD}=\frac{m}{2}\frac{f_{(AB}f_{CD)}}{S} \ .}
	\end{equation}
Below we will show that this relationship is satisfied for Type N spacetimes with a scalar field $S$ satisfying the massive Klein Gordon equation with a non-minimal coupling in curved spacetimes. The scalar $S$ can be thought of as a linearised solution of the massive biadjoint scalar when considering the ansatz $S^{ab}=c^ac^b S$; this is commonly referred to as the zeroth copy.

\section{Type N solutions}
For Type N solutions, the Cotton spinor and field strength spinor can be written as
	\begin{equation} \label{eq:cotton dc spinor}
	  C_{ABCD}=\psi_4 o_A o_B o_C o_D \ , \quad    f_{AB}=\Phi_2 o_A o_B \ ,
	\end{equation}
where $\psi_4$ and $\Phi_2$ are Newman-Penrose (NP) scalars. In this case, the double copy can simply be expressed as 
    \begin{equation}
    \psi_4=\frac{m\Phi_2^2}{2 S} \ . \label{eq:CottonDCNP}
    \end{equation}
We will now prove that the Cotton double copy holds for type N spacetimes in curved backgrounds by deriving the equation of motion of the zeroth copy $S$.

We start by substituting \eqref{eq:cotton dc spinor} into \eqref{eq:C eom} and \eqref{eq:f eom}, and contracting the equations with $\iota$ and $o$ to get:
\begin{align} 
o_{A} {\nabla^{A}}_C \log \Psi_{4}+4 o_{A} \iota^{B} {\nabla^{A}}_C o_{B}-\iota_{A} o^{B} {\nabla^{A}}_C o_{B}=\frac{m}{\sqrt{2}}o_c \ , \label{eq:psi4}\\
o_{A} {\nabla^{A}}_C \log \Phi_{2}+2 o_{A} \iota^{B} {\nabla^{A}}_C o_{B}-\iota_{A} o^{B} {\nabla^{A}}_C o_{B}=\frac{m}{\sqrt{2}}o_c \ ,  \label{eq:phi2} \\
o^B o_C\nabla^{CA}o_B=0 \ . \label{eq:OOO}
\end{align}
From the Cotton double copy in \eqref{eq:CottonDCNP}, together with Eqs.~\eqref{eq:psi4} and \eqref{eq:phi2}, we find
\begin{equation} \label{eq:eom S}
 o_{A} {\nabla^{A}}_C \log S-\iota_{A}{o}^{B} {\nabla^{A}}_C{o_{B}}=-\frac{m}{\sqrt{2}}o_c \ .
\end{equation}
To show that $S$ satisfies the Klein-Gordon equation with a non-minimal coupling term first we write $\nabla_{\mu}\nabla^{\mu} S$ as 
\be
-\nabla_{AB}\nabla^{AB} S=-\epsilon_{AC}{\nabla^C}_{B}\nabla^{AB} S=2\iota_{C}o_{A}{\nabla^C}_{B}\nabla^{AB} S.
\ee
Then, one can use the Leibniz rule and \eqref{eq:eom S} to eliminate the derivatives of $S$:
\begin{equation}
\begin{split}    
 \label{eq:box S}
2\iota_{C}o_{A}{\nabla^C}_{B}\nabla^{AB} &S=-2\iota_{C}{\nabla^{C}}_B o_{A}\nabla^{AB} S\\&+2\iota_{C}{\nabla^C}_{B}\left(S\iota_{A}{o}^{D} {\nabla^{AB}}{o_{D}}-\frac{m}{\sqrt{2}}S o^B\right).
\end{split}
\end{equation}
Expanding \eqref{eq:box S} and using \eqref{eq:eom S} to eliminate $\nabla S$ terms we get:
\be \label{eq: S eom all terms}
\begin{split}
&-\nabla_{AB}\nabla^{AB} S= S \bigg(2\iota_{C} {\nabla^C}_{B}\iota_{A}{o}^{D} {\nabla^{AB}}{o_{D}} \\
&+2\iota_{C}\iota_{A}{\nabla^C}_{B}{o}^{D} {\nabla^{AB}}{o_{D}}+2\iota_D\iota_{A}{\nabla^{A}}_C{o^{D}}\iota_{E}{o}^{F} {\nabla^{EC}}{o_{F}}\\&-m\sqrt{2}\left(\iota_{C} {\nabla^C}_{B}o^B+\iota_D\iota_{A}{\nabla^{A}}_C{o^{D}} o^C-\iota_{C}\iota_{E}{o}^{F} {\nabla^{EC}}{o_{F}}\right) \\
  &+m^2+2\iota_{C}\iota_{A}{o}^{D} {\nabla^C}_{B}{\nabla^{AB}}{o_{D}}\bigg) \ .
\end{split}
\ee
The first three terms as well as the terms linear in $m$ add up to zero by \eqref{eq:OOO}. The term with the second derivative of $o$ can be related to curvature spinors by the following relation \cite{castillo20033}:
\be
{\nabla}_{D(A}{\nabla_{B)}}^D o_C=\frac{1}{2}\Phi_{ABCD}o^D+\frac{1}{24}R\left(\epsilon_{AC}o_B+\epsilon_{BC}o_A\right),
\ee
where $\Phi_{ABCD}$ is the spinor equivalent of traceless Ricci tensor which is proportional to Cotton spinor by TMG equation of motion. By substituting \eqref{eq:cotton dc spinor}, we see that the term proportional to $\Phi_{ABCD}$ does not contribute. Therefore we find that
\begin{align*}
&2\iota_{C}\iota_{A}{o}^{D} {\nabla^C}_{B}{\nabla^{AB}}{o_{D}}=\frac{1}{6}R \ . 
\end{align*}
Finally, substituting everything into \eqref{eq: S eom all terms} we get
\be
-{\nabla}_{AB}\nabla^{AB} S=\Box S=\left(m^2+\frac{1}{6}R\right)S. \label{eq:scalarEOM}
\ee
This proves that the Cotton double copy is satisfied for Type N solutions with the zeroth copy given by a linearized massive bi-adjoint scalar with a non-minimal coupling. Note that we obtained the same non-minimal coupling as in the 4d zeroth copy \cite{Carrillo-Gonzalez:2017iyj,Bahjat_Abbas_2017,White:2020sfn}, but in 3d it does not give a conformally invariant equation. We will now show explicit examples of Type N spacetimes where the Cotton double copy holds.

\subsection{pp-waves}
In this subsection, we analyze the double copy relation for plane-fronted waves with parallel propagation (pp-waves). For TMG, any solution that admits a null Killing vector, well-defined through all space, is a pp-wave solution \footnote{The nomenclature of pp-waves for the non-zero cosmological constant case can be misleading since the null Killing vector is not covariantly conserved \cite{garcia-diaz_2017}.} 
\cite{Gibbons:2008vi}. In flat space the metric of pp-waves can always be written as \cite{Chow:2009km}
    \begin{equation}
    \mathrm{d} s^{2}=\mathrm{d} y^{2}-2 \mathrm{~d} u \mathrm{~d} v+\mathrm{e}^{-m y} f(u)\mathrm{d} u^{2} .
    \end{equation}
while in AdS it reads
	\begin{equation}
    \mathrm{d} s^{2}=\mathrm{d} y^{2}-2 \mathrm{e}^{2 \frac{y}{L}} \mathrm{d} u \mathrm{~d} v+\mathrm{e}^{\frac{(1-mL)}{L} y} f(u) \mathrm{d} u^{2} ,
    \end{equation}
where $L$ is the AdS radius. Note that we can obtain the dS solution by taking $L\rightarrow i L$. On the TMYM side, we can write the pp-wave solution as 
 \be
A^a=c^a\mathrm{e}^{-m y} g(u) du,
\ee
for the Minkowski, AdS, and dS cases. In Table~\ref{table:pp-waves} we show the NP scalars for the corresponding pp-waves in Minkowski and AdS. One can easily see that the scalar $S$, which is computed using the Cotton double copy in Eq.~\eqref{eq:CottonDCNP} satisfies
\be
(\nabla^2-m^2-\frac{R}{6})S(u,y)=(\partial_y^2-m^2+\frac{1}{L^2})S=0 \ .
\ee

\begin{table}[!h]
\begin{tabular}{c|c|c|c|}
\cline{2-4}
                                & $\Psi_4$ & $\Phi_2$ & $S$ \\ \hline
\multicolumn{1}{|l|}{Mink.} &  $\frac{m^3}{4} \mathrm{e}^{-m y}f(u) $ &  $-\frac{m}{\sqrt{2}} \mathrm{e}^{-m y} g(u)$         &  $\frac{g(u)^2}{f(u)}\mathrm{e}^{-m y}$   \\ \hline
\multicolumn{1}{|l|}{AdS}     &  $\frac{m^3}{4}\mathrm{e}^{-(\frac{3}{L}+m)y}f(u)$        & $-\frac{m}{\sqrt{2}} \mathrm{e}^{-(\frac{2}{L}+m)y} g(u)$         & $\frac{g(u)^2}{f(u)}\mathrm{e}^{-(\frac{1}{L}-m) y}$     \\ \hline
\end{tabular}
\caption{In this table we show the NP-scalars for the Cotton spinor and the dual field strength spinor. We also show the scalar $S$ constructed from the Cotton double copy in Eq.~\eqref{eq:CottonDCNP}.}
\label{table:pp-waves}
\end{table}

\subsection{Shock waves and gyratons}
\subsubsection{Minkowski}
We will now consider solutions with a source corresponding to a fast moving particle whose stress tensor is traceless and is given by 
	\be
	T_{\mu\nu}=\left(E k_{\mu}k_{\nu}+  \sigma k_{(\mu }\epsilon_{\nu)}^{\alpha \beta} k_{\alpha} \partial_{\beta}\right)\delta(u)\delta(y) \ ,
    \ee	
    where the null vector $k^\mu$ is defined as $k_\mu dx^\mu= du $, E is the energy of the source particle and $\sigma$ is its classical spin. Note that this source can be thought of as boosted gravitational anyon. If the particle has no classical spin ($\sigma=0$) then it generates shockwaves; otherwise, the solutions are dubbed gyratons. In flat space, both of these solutions have a metric of the form \footnote{Note that the gyraton metric is generically written as $ds^2=-2du dv+dy^2+\kappa F(u,y) du^2 +2\kappa \alpha(u,y)du dy$, where the cross term proportional to $\alpha(u,y)$ allows us to see the rotation explicitly \cite{Frolov:2005zq}. Here we have chosen a gauge where $\alpha=0$.}
    \begin{equation}
    \mathrm{d} s^{2}=\mathrm{d}y^{2}-2 \mathrm{~d} u \mathrm{~d} v+\kappa F(u,y)\mathrm{d} u^{2} .
    \end{equation}
For these solutions, we have that the only non-zero NP Cotton scalar is
	\begin{equation}
	   \psi_4=-\frac{1}{4}\partial_y^3F(u,y) \ ,
	\end{equation}
where $F$ satisfies the following equation of motion
	\begin{equation}
	\partial_y^3F(u,y)+m \partial_y^2F(u,y)=\kappa  m \delta(u)  \left(E\delta(y) -\sigma \delta'(y) \right) \ .
	\label{eq:eomTMG}
	\end{equation}
On the gauge theory side, we will also consider a boosted spinning source whose current is given by
    \be 
	 J_{\mu}=\left(Q k_{\mu}+  Q' \epsilon_{\mu}^{\alpha \beta} k_{\alpha} \partial_{\beta}\right)\delta(u)\delta(y) \ ,
	 \ee
where $Q$ is the electric charge and $Q'$ contributes, together with $Q$, to the magnetic flux. We consider the following gauge field
    \be
	 A^a=c^a G(u,y) du, \label{eq:TMEwave}
	 \ee
which linearises the TMYM equations of motion and gives only one non-vanishing component of field strength $F_{uy}=-\partial_y G(u,y)$. Hence the only non-zero NP field strength scalar is
	\begin{equation}
    \Phi_2=\frac{1}{\sqrt{2}}\partial_y G(u,y) \ ,
	\end{equation}
where $G$ satisfies
    \be
    \partial_y^2 G(u,y) + m \partial_y G(u,y) =g \delta(u) \left(Q \delta(y)-Q' \delta'(y)\right) \ . \label{eq:eomTMYM}
    \ee
Then the scalar $S$ in the Cotton double copy, Eq.~\eqref{eq:cottonDCspinor}, is given as
	 \be
	 S=-m\frac{(\partial_y G(u,y))^2}{\partial_y^3 F(u,y)}.
	 \ee
Equations \eqref{eq:eomTMG} and \eqref{eq:eomTMYM} imply that outside the sources the following is true:
	 \be
	 (\nabla^2-m^2)S(u,y)=0.
	 \ee

To see the double copy for an explicit gyraton or shock wave solution, we need to pick boundary conditions for the metric which is equivalent to picking the $i \epsilon$ prescription when regulating the phase shift felt by a particle moving on the shockwave background. As realised in \cite{Deser:1993wt}, we cannot have the same coordinate chart on both sides of the shockwave. In \cite{Gonzalez:2021ztm}, we have shown that a useful prescription to observe the double copy relation is to consider boundary conditions where the metric is Cartesian on one side and flat in the other side \cite{Edelstein:2016nml}. For example, in the gravitational shock wave case 
\begin{equation}
    F(u,y)=-\kappa\frac{ E }{m } \left(  e^{-m y }  \theta(y) +  (1-m y) \theta(-y)\right)\delta(u) \ .
\end{equation}
Choosing the analogue boundary condition in TMYM leads to 
\begin{equation} \label{eq: tmym shock wave Cartesian}
		G(u,y)=\frac{gQ}{m}\left(e^{-my}\theta(y)+\theta(-y)\right)\delta(u) \ .
\end{equation}
On the $y<0$ side of the shock wave, the double copy is trivial since $\Psi_4=\Phi_2=0$. On the other hand, in the $y>0$ side the NP scalars are proportional to those of the flat space pp-waves in Table \ref{table:pp-waves} times $\delta(u)$. Making the replacement $Q^2\rightarrow E $ leads to the double copy relation in Eq.~\eqref{eq:CottonDCNP}. Note that the gyratons double copy can be constructed analogously, and is related to the shockwave solution by the shift $ E \rightarrow E\left(1+m\frac{\sigma}{E} \right)$ in TMG and $Q\rightarrow Q(1+m\frac{Q'}{Q})$ in TMYM, which arise from spin deformations of on-shell 3-point amplitudes \cite{Burger:2021wss} and was originally found for gravitational anyons \cite{PhysRevLett.64.611}.

\subsubsection{AdS}
We proceed to consider the case of shock wave solutions of TMG, TMYM in an AdS background. We start with the gravitational shockwaves in AdS which can be written in Poincare coordinates as 
    \begin{equation}
	ds^2=\frac{L^2}{y^2}\left(-2du dv+dy^2 +\delta(u)F(y)du^2 \right)\ ,
		\label{eq:TMGshockAdS}
	\end{equation}
	where $F$ satisfies:
	\begin{equation}
	 \frac{y}{L}F''' +m F''-m \frac{F'}{y}  = \kappa E \frac{L}{y} m \delta(y-y_0) \ ,
	\end{equation}
where $y_0\neq 0$ is the location of the source in the bulk and we will assume $mL>1$. As before, we need to fix the boundary conditions to find the explicit solution. We do so by choosing to have the same boundary conditions as in the Minkowski case in the flat space limit. This is equivalent to imposing Brown-Henneaux boundary conditions and requiring a regular solution in the bulk. The explicit solution with these boundary conditions is
	\begin{equation}
	\begin{split}
	    F(y)=&-\frac{\kappa L^{2} m E}{2\left(1-(L m)^{2}\right)}\Bigg[2\left(\frac{y}{y_{0}}\right)^{1-L m}\theta\left(y-y_{0}\right)\\
	    &+\left((1-L m)\left(\frac{y}{y_{0}}\right)^{2}+(1+L m)\right)\theta\left(-y+y_{0}\right)\Bigg]  \ .
	\end{split}
	\end{equation}
On the non-trivial side of the shock wave, $y>y_0$, we have that the only non-zero Cotton NP scalar is 
	\begin{equation}
	    \Psi_4=-\frac{1}{2}\frac{y}{L}\delta(u)F'''(y) =\frac{\kappa}{2}E L^2 m^2 \left(\frac{y}{y_{0}}\right)^{-1-L m}  \delta(u) \ .
	\end{equation}

On the other hand, the linearised shockwave solution for TMYM in an AdS background is given by 
    \be
	 A^a=c^a \delta(u)G(y) du, 
	 \ee
where the function $G$ satisfies 
	\begin{equation}
	  \frac{y^4}{L^4}\left( G''+\frac{1+ L m}{y} G' \right)= \frac{y^3}{L^3}g Q \delta(y-y_0) \ .
	\end{equation}
The explicit shockwave solution with boundary condition analogue to the gravitational case is given by 
		\begin{equation}
	    G(y)=-\frac{g Q}{m}\left[\left(\frac{y}{y_0}\right)^{-L m}\theta\left(y-y_{0}\right)-\theta\left(-y+y_{0}\right) \right] \ .
	\end{equation}
Thus we have that the only non-zero dual field strength NP scalar is
	\begin{equation}
    \Phi_2=2 \delta(u)\frac{y}{L}f'(y)=2 g Q\delta(u) \left(\frac{y}{y_0}\right)^{- L m}\ .
	\end{equation}
	
Lastly, we consider the linearised biadjoint scalar, $S^{a\tilde{a}}=c^a c^{\tilde{a}} S$, living in an AdS background with a non-minimal coupling as in Eq.~\eqref{eq:scalarEOM}. 
The shockwave solution is now given by 
    \begin{equation}
    \begin{split}
	 S=&\frac{1}{2 m}\left(\frac{y_0}{L}\right)^2\Bigg[\left(\frac{y}{y_0}\right)^{1-L m}\theta\left(y-y_{0}\right) \\ 
	 &-\left(\frac{y_0}{y}\right)^{-1-L m}\theta\left(-y+y_{0}\right)\Bigg]\delta(u) \ ,
	 \end{split}
	\end{equation}
where we chose boundary conditions by requiring that the field vanishes deep in the bulk and as we approach the AdS boundary. We note that this solution corresponds to the scalar that arises from the Cotton double copy relation in Eq.~\eqref{eq:CottonDCNP}, which shows that this relation is satisfied as expected for AdS shock waves. In a similar manner to the flat space case, one can consider shifts of the charge and energy to obtain the gyraton double copy. In this case the shifts are given $ E \rightarrow E\left(1+\frac{\sigma}{E}\left((1+ mL)/y_0\right) \right)$ for the TMG case and $Q\rightarrow Q(1+\frac{Q'}{Q}\left((1+ mL)/y_0\right))$ in TMYM. In future work, we will explore whether the origin of these shifts can be traced down to spin deformations of three-point correlators.
	
\section{Conclusions and future directions}
We have shown that it is possible to construct a double copy relation for topologically massive theories that gives the Cotton spinor as the square of the dual field strength spinor in curved spacetime backgrounds. This generalises the 4d Weyl double copy to 3d spacetimes. In this letter, we have focused on Type N spacetimes. We gave a proof of the Cotton double copy for gravitational waves and showed explicit examples. Other examples that we didn't look at explicitly can be found in \cite{garcia-diaz_2017,Chow:2009vt}. It would be interesting to understand whether the double copy holds for Type D solutions and for solutions with sources that have a non-zero trace of the stress-energy tensor. Previous analysis \cite{Gonzalez:2021ztm,Burger:2021wss,Moynihan:2021rwh} have shown that this is not straightforward, and further investigations should clarify this intriguing case.

\section{Acknowledgments}
We would like to thank Chris D. White for his insightful comments on our letter and Nathan Moynihan for useful discussions. MCG is supported by the European Union’s Horizon 2020 Research Council grant 724659 MassiveCosmo ERC–2016–COG and the STFC grants ST/P000762/1 and ST/T000791/1. JR is supported by an STFC studentship. AM is partially supported by the \emph{Séjours Scientifiques de Haut Niveau} fellowship. AM thanks \emph{L'Institut de Physique Théorique} and \emph{L'Institut des Hautes Études Scientifiques} for  their hospitality during the early stage of this work.

\paragraph{Note added}
During the completion of this letter, we were made aware of the work by N. Moynihan and W. Emond which contains some overlapping results.

%%%%%%%%%%%%%%%%
\bibliographystyle{JHEP}
\bibliography{references}
%%%%%%%%%%%%%%%%

\end{document}